\documentstyle[12pt,supercite]{article}
\newcommand{\Obs}{{\cal O}}

\newcommand{\la}{\langle}
\newcommand{\ra}{\rangle}

\newlength{\digitwidth} 

\def\sqr#1#2{{\vcenter{\hrule height.#2pt
      \hbox{\vrule width.#2pt height#1pt \kern#1pt
        \vrule width.#2pt}
      \hrule height.#2pt}}}

\def\abstracts#1#2#3{{
        \centering{\begin{minipage}{4.62in}\baselineskip=13pt
        \small
        \centerline{\bf Abstract}
        \vspace*{0.2cm}                
        \parindent=0pt #1\par
        \parindent=18pt #2\par
        \parindent=15pt #3
        \end{minipage} }\par}}

\renewcommand{\thefootnote}{\fnsymbol{footnote}}
\begin{document}
\vspace*{-3cm}
\hfill \parbox{4.2cm}{ KOMA-96-15\\
                       FUB-HEP 07/96\\
                       May 1996\\
               }\\
\vspace*{1.3cm}
\centerline{\LARGE \bf Multigrid Method } \\[0.5cm]
\centerline{\LARGE \bf versus } \\[0.5cm]
\centerline{\LARGE \bf Staging Algorithm } \\[0.5cm]
\centerline{\LARGE \bf for PIMC Simulations } \\[1.5cm]
\renewcommand{\thefootnote}{\arabic{footnote}}
\vspace*{0.3cm}
\centerline{\large {\em Wolfhard Janke\/}$^{1}$ 
               and {\em Tilman Sauer\/}$^2$}\\[0.4cm] 
\centerline{\large    $^1$ {\small Institut f\"ur Physik,
                      Johannes Gutenberg-Universit\"at Mainz}}
\centerline{    {\small 55099 Mainz, Germany }}\\[0.15cm]
\centerline{\large    $^2$ {\small Institut f\"{u}r Theoretische Physik,
                      Freie Universit\"{a}t Berlin}}
\centerline{    {\small 14195 Berlin, Germany}}\\[2.50cm]
\vspace*{0.3cm}
\abstracts{}{
We present a comparison of the performance of two non-local update algorithms
for path integral Monte Carlo (PIMC) simulations,
the multigrid Monte Carlo method and the staging algorithm.
Looking at autocorrelation times for the internal energy we show that
both refined algorithms beat the slowing down which is encountered for
standard local update schemes in the continuum limit. We investigate the
conditions under which the staging algorithm performs optimally and
give a brief discussion of the mutual merits of the two algorithms.
}{}
\vspace*{0.5cm}
  \thispagestyle{empty}
      \newpage
       \pagenumbering{arabic}
%
%
%
       \section{Introduction}\label{sect:Introduction}
%
A well-known problem for path integral Monte Carlo simulations \cite{slow_pimc}
using standard local update schemes such as the Metropolis algorithm
is a severe slowing down in the continuum limit. By this one means that
successively generated configurations in the Monte Carlo process
are highly correlated, a phenomenon signalized by large autocorrelation
times in the simulation.
This slowing down problem is very similar to the critical slowing down
encountered in simulations of statistical or lattice field theoretical
systems near phase transitions of second order \cite{csd}.
In both cases it is the diverging spatial correlations (in lattice units)
which are the physical origin of the inefficiency
of local update schemes.

In many of the applications
in statistical physics and lattice field theory
the critical slowing down problem can be 
overcome by the use of multigrid techniques \cite{multigrid}.
These are non-local 
update schemes where updates are performed on a variety of length
scales in order to sample most efficiently long wave-length fluctuations.
In a recent letter \cite{js93a} we have shown that 
thanks to the generality of their definition multigrid techniques
can be transferred to simulations of Euclidean path integrals.
We explicitly demonstrated that also for these systems
slowing down is almost completely 
reduced. Another advantage of their general definition is that
multigrid techniques can also be combined with reweighting schemes such as 
multicanonical sampling in order to further reduce autocorrelation times
in the presence of tunneling barriers \cite{js94,js95b}.

For Monte Carlo simulations of path integrals another successful
non-local update algorithm, which is somewhat similar in spirit
but which works technically in a rather different manner,
is known under the name of ``staging'' \cite{staging1,staging2,tuckerman}.
Although this algorithm also significantly reduces the slowing down of
simulations in the continuum limit, to our knowledge no
detailed analysis of autocorrelation times for the staging
algorithm exists in the literature. 
Also, we know of no work which gives a comparison of the two algorithms.
In this letter, we intend to fill this gap by reporting autocorrelation
times for a standard energy estimator and two sample potentials, a convex
one and a double well,
employing both the staging algorithm and the multigrid method.
%
       \section{The Algorithms}\label{sect:The Algorithms}
%
In the path integral Monte Carlo approach the
quantum partition function ${\cal Z}$ at inverse
temperature $\beta$ is approximated as a discretized path integral consisting
of $L$ ``beads'' \cite{kleinert90}
\begin{equation}
     {\cal Z}_L(\beta) =
              \left[ \prod_{i=1}^{L}  \int \frac{dx_i}{A} \right]
     \exp\{- {\cal A}_L \},
 \label{DiscretePI}
\end{equation}
with an action
\begin{equation}
         {\cal A}_L =  \epsilon \sum_{i=1}^{L}
         \left[\frac{1}{2}(\frac{x_i-x_{i-1}}{\epsilon})^2 + V(x_i)\right],
 \label{Action}
\end{equation}
where $V$ is a potential to be specified below,
$A=\sqrt{2\pi\epsilon}$, $\epsilon=\beta/L$, $x_0 = x_L$, and
$\hbar = k_B = 1$.
The original partition function ${\cal Z}$ is then recovered in the continuum
limit $L\rightarrow\infty$ with $L\epsilon = \beta$ fixed.

The basic idea of the multigrid approach \cite{multigrid} is to perform 
non-local
updates of the $x_i$ by working on a set of successively coarser discretizations
of the time axis (''grids``) in order to take
into account long wave-length fluctuations of the paths more efficiently. The technical 
details of this algorithm have been described in detail elsewhere, see e.g. 
Ref.~\cite{js93a}. The algorithm requires the definition of a set of coarser 
grids and a prescription to set up coarsened actions on these grids. 
Given the grids and the corresponding actions the 
multigrid algorithm recursively defines a sequence in which the variables on the
various grids are updated and interpolated back onto the original grid.
Two of those sequences which have extensively been used 
and studied are known under the name of V-cycle and W-cycle.

The basic idea of the staging method is to rewrite the quantum statistical 
partition function in such a way that a sequence of $j$ adjacent variables 
can be 
updated independently. Technical details can again be 
found in the literature \cite{staging1,staging2,tuckerman}. The algorithm 
implies the random selection of 
the end points of some segment of the path of length $j$ and to perform a 
change of variables which allows an elimination of the nearest neighbor 
coupling stemming from the kinetic energy. For the variables of the staging 
segment the partition function hence reduces to a collection of independent
oscillators moving in an external potential which depends on the 
transformation of the variables. The staging variables may then be updated 
using Gaussian random numbers and a Metropolis like acceptance rule for the
external potential. In contrast to the multigrid scheme the staging algorithm
only allows for one single tunable parameter,
namely the length $j$ of the staging segment.
%
       \section{Results}\label{sect:Results}
%
As in Ref.~\cite{js93a} we studied two qualitatively different potential 
shapes, typical for a wide range of physical phenomena. A convex potential (CP),
given by $V = 0.5x^2+x^4$, is relevant for studying fluctuations
around a unique minimum, and a double-well potential (DW), given by
$V = -0.5x^2+0.04x^4$, is relevant for studying tunneling phenomena.
We have simulated the path integral (\ref{DiscretePI}), (\ref{Action}) 
for grids of size $L = 2^3=8$ up to 
$2^{10}=1024$, using both the multigrid scheme with Metropolis update and 
the staging algorithm. In all our simulations $\beta$ was equal to $10$.

An observable of central importance is the internal energy. For path integral
Monte Carlo simulations two different estimators for this observable are 
well-known \cite{barker79,cf81,hbb82}. 
Straightforward application of the definition
of the internal energy $U_L = -\partial \ln {\cal Z}_L/\partial\beta$ leads to an 
estimator $U_L = U_L^{\rm kin} + \frac{1}{L} \sum_{i=1}^L \la V(x_i) \ra$ with
$U_L^{\rm kin} = L/2\beta - \frac{1}{L} \sum_{i=1}^L \la \frac{1}{2}
\left(\frac{x_i - x_{i-1}}{\epsilon} \right)^2 \ra$.
We call $U_L$ the ``kinetic'' estimator since it explicitly measures
the kinetic part $U_L^{\rm kin}$ of
the energy \cite{barker79}. Another function of the Monte Carlo configurations
is given by $\frac{1}{L} \sum_{i=1}^L \left[ \frac{1}{2} x_i V'(x_i) + V(x_i)
\right]$,
and its expectation value $\la \dots \ra$ with respect to (\ref{DiscretePI}) 
also estimates the internal energy \cite{cf81,hbb82}.
We call it the ``virial'' estimator since it can be obtained by invoking
the virial theorem. 
As usual expectation values are approximated in the simulation by mean values
of the estimators over the Monte Carlo sequence.
For update schemes which reduce slowing down in the 
continuum limit the virial estimator is a priori superior in this limit since
it has a constant variance whereas the variance of the kinetic estimators 
grows linearly with $L$. Clearly, the best estimator would be some linear
combination of the two estimators which must take into account the
individual variances {\em and} the covariance of the two
estimators \cite{js95b,js95a}.
For our comparison of multigrid and staging update schemes we will
here, however, only look at the virial estimator.
A careful discussion of the problem of optimized energy estimation
in path integral Monte Carlo simulations is beyond the scope of this letter
\cite{jstobepub}.

Since the main focus of this letter is to present a comparison of the
performance of the two different non-local update schemes we have taken care
to measure precisely the autocorrelation times obtained 
for the multigrid and staging algorithms.
Explicitly, the autocorrelation function $A(k)$ of an observable $\Obs$ is 
defined by \cite{ms88}
\begin{equation}
A(k) = \frac{\la \Obs_i \Obs_{i+k}\ra - \la \Obs_i \ra^2}
            {\la \Obs_i^2 \ra - \la \Obs_i \ra^2} ,
\end{equation}
where $\Obs_i$ stands short for the $i$th measurement of $\Obs$.
The autocorrelation time $\tau_0$ then follows from the asymptotic behaviour
for large $k$, $A(k) \propto \exp(-k/\tau_0)$. The integrated autocorrelation
time $\tau$, defined by the area under the autocorrelation function of this 
observable,
\begin{equation}
\tau \equiv \frac{1}{2} + \sum_{k=1}^{\infty} A(k),
\label{Tau_int}
\end{equation}
usually behaves qualitatively as
$\tau_0$. It can be shown to enter in the estimate for the statistical
error of mean values as 
$\Delta\hat{\Obs}=\sqrt{2\tau}\sqrt{\sigma^2/N_m}$, where $\sigma^2$ is the
observable's variance and $N_m$ the number of measurements used to compute 
the mean value $\hat{\Obs}$. The effective statistics is thus reduced to 
$N_{\rm eff}=N_m/2\tau$. Since $A(k)$ becomes very noisy for large $k$, the
upper bound in (\ref{Tau_int}) is usually cut off self-consistently at
$f \tau$ with $f \approx 6 \dots 8$ \cite{js95b,ms88}.
In our analysis we used $f=8$, and all error bars for these
autocorrelation times were obtained by jackkniving \cite{jack}
the data with $100$ blocks.

In our simulations we performed $N_m = 100\,000$ sweeps 
after discarding $5\,000$ sweeps for thermalization.
Measurements were taken after each sweep.
``Sweep'' here means a complete V- resp. W-cycle in the case of
the multigrid algorithm and ${\rm int}(L/(j-1))$ calls to the staging routine
which moves $j-1$ adjacent variables at each call. 
Notice that the above definition of a staging sweep in general implies updates
of less than $L$ variables. We have therefore rescaled the measured 
autocorrelation times by a factor $({\rm int}(L/(j-1)))/(L/(j-1))$.

For the staging algorithm the length $j$ of the segment which is to
be updated in the staging routine is the only
parameter which may be tuned in order to optimize the performance.
A rule of thumb here says that it should be set to such a value that
the acceptance rate is $40\%$ \cite{tuckerman}.
For the above definition of a staging pass the amount of
numerical work to be done does not depend significantly on the parameter $j$.
The only criterion for the optimal choice of $j_{\rm opt}$ therefore is the
integrated autocorrelation time for the observable at hand.
Figure 1a shows
the integrated autocorrelation times 
as a function of $j$ for different values of $L$
using the virial estimator for the internal energy of the convex potential.
We notice that there certainly is an optimal value $j_{\rm opt}$ 
for each $L$ even though the minima are quite shallow and
autocorrelation times do not differ very much for neighbouring values of
$j_{\rm opt}$. 
We also see that the autocorrelation times for $j_{\rm opt}$ stay roughly
constant, independent of $L$. It is thus already seen qualitatively that
the staging algorithm eliminates slowing down completely.

As discussed above, it is the divergence of the correlation length
measured in lattice units which is the cause of the slowing down problem in
the continuum limit. The staging algorithm
overcomes this problem by updating a whole segment of the path in a
completely decorrelating manner.
One would hence expect that the efficiency of the staging algorithm
depends on the ratio of the length of the staging segment to the
correlation length along the path. Since the latter scales with $L$
one would therefore expect that the optimal choice of $j$ should also
scale with $L$. In Fig.~1b we have therefore plotted the autocorrelation times 
with a rescaling of the $x$-axis to $j/L$. Except for small values of 
$L$ we notice that the curves do indeed collapse onto a common master 
curve under such a rescaling. Figure 1b thus shows in particular that 
the optimal value $j_{\rm opt}$ scales with $L$. 
This observation shows that the optimal staging parameter 
$j_{\rm opt}$ can in practice easily be obtained by looking at the 
autocorrelation times for moderately coarse discretization. The optimal 
choices of $j$ can then be obtained for any finer discretization 
by a simple rescaling.

For the virial estimator the rule of $40\%$ acceptance probability turns out
not to be too far off the mark. For the optimal $j_{\rm opt}$ we
find an acceptance rate of roughly $55\%$ independent of the grid size
$L$ for not too small grids (cp.~Table 1). 
\begin{table}[tb]
\catcode`?=\active \def?{\kern\digitwidth}
\centering
\caption{Integrated autocorrelation times for the
         virial estimator of $U$ using 
         the staging algorithm ($\tau_{\rm s}$) and
         the multigrid algorithm
         with $V$-cycle ($\tau_{\rm v}$) and W-cycle
         ($\tau_{\rm w}$). 
         The second and third columns give the optimal segment length and the 
         acceptance rate in percent for the staging
         algorithm.
         }
\vspace*{0.3cm}
\begin{tabular}{|r|r|l|l|l|l|}
\hline
\multicolumn{1}{|c}{$L$}              &
\multicolumn{1}{|c}{$j_{\rm opt}$}    & 
\multicolumn{1}{|c}{\%}               &
\multicolumn{1}{|c}{$\tau_{\rm s}$}   &
\multicolumn{1}{|c}{$\tau_{\rm v}$}   &
\multicolumn{1}{|c|}{$\tau_{\rm w}$}  \\
\hline
\multicolumn{6}{|c|}{Convex Potential (CP)}\\
\hline
8    &   2 & 64 & 1.545(32) & ?1.033(16) & 0.851(16)   \\
16   &   2 & 82 & 1.518(24) & ?0.912(14) & 0.692(12    \\
32   &   4 & 72 & 1.850(44) & ?0.909(18) & 0.644(11)   \\
64   &  10 & 68 & 1.959(44) & ?1.092(22) & 0.6278(66)  \\
128  &  24 & 55 & 2.012(49) & ?1.583(31) & 0.660(14)   \\
256  &  44 & 55 & 2.119(67) & ?2.701(82) & 0.7119(93)   \\
512  &  88 & 55 & 2.079(53) & ?5.07(21)  & 0.781(12)   \\
1024 & 176 & 56 & 2.048(56) & 10.80(94)  & 0.859(16)   \\
\hline
\multicolumn{6}{|c|}{Double-Well Potential (DW)}\\
\hline
8    &   2 & 71 & 1.779(36) & ?0.752(18) & 0.5478(76) \\
16   &   6 & 47 & 2.059(44) & ?0.746(12) & 0.5121(79) \\
32   &  10 & 49 & 2.182(63) & ?0.784(12) & 0.5128(79) \\
64   &  20 & 52 & 2.335(64) & ?0.945(16) & 0.5291(75) \\
128  &  40 & 54 & 2.406(76) & ?1.338(25) & 0.5377(72) \\ 
256  &  80 & 54 & 2.456(87) & ?2.297(61) & 0.5479(72) \\
512  & 160 & 54 & 2.366(64) & ?3.92(15)  & 0.5650(95) \\
1024 & 320 & 54 & 2.406(69) & ?7.19(43)  & 0.5920(76) \\
\hline
\end{tabular}
\label{table:2ukuv}
\end{table}
To give a numerical example, for the convex potential and a medium sized 
grid of $L=256$ the optimal value is $j_{\rm opt} = 44$.
An acceptance of $40\%$ on the other hand would be achieved for 
$j_{40\%} = 72$.
For $j_{40\%}$ we then find an integrated autocorrelation time of $2.662(96)$ 
which is significantly larger than the minimal value of $2.119(67)$ 
(cp.~Table 1). We also emphasize that the integrated autocorrelation times 
do depend on the observable and on its estimator. Thus the $40\%$ rule may 
be rather misleading for a different observable resp. estimator. In fact,
we found that the acceptance probabilities in the staging algorithm for that 
$j_{\rm opt, k}$ which optimizes the autocorrelation times for the kinetic 
estimator is roughly $90\%$ \cite{jstobepub}. For our comparison with the 
multigrid update schemes we have in any case used the value $j_{\rm opt}$
which minimized the integrated autocorrelation time.

Let us now turn to the comparison of the two update schemes. 
For the multigrid algorithm
autocorrelation times for the moments $\la x^n \ra$, $n=1,\dots,4$
were already reported in Ref.~\cite{js93a}. Here we look at the virial 
estimator for the internal energy but in order to facilitate comparison 
with our previous results we have used the same parameters as in 
Ref.~\cite{js93a} for the multigrid scheme.
This means in particular that we performed
only presweeps and no postsweeps \cite{js93a}. 
The acceptance rates for the finest grid
were adjusted to be roughly $50\%$.
Table 1 lists for the virial estimator the integrated autocorrelation times
$\tau$
for the V-cycle and the W-cycle
as well as
for the staging algorithm with $j_{\rm opt}$ as discussed above.
These data are also plotted in Figs.~2a and 2b,
together with fits to the data according to a 
power law of the form $\tau = \alpha L^z$. For the V-cycle these fits were 
done on the basis of the data for the three largest grids, for the W-cycle 
and the staging algorithm data for the four largest grids were used for the 
fits.

Since for a polynomial potential the virial estimator is a linear combination
of the expectation values for the moments $\la x^n \ra$ one would expect that the 
autocorrelation times for the virial estimator would not differ too much from
the autocorrelation times for the moments. A comparison with the data reported
in Ref.~\cite{js93a} shows indeed a qualitative agreement. The multigrid 
V-cycle again reduces the $L^2$ divergence of autocorrelation times for 
standard local updates to a linear dependence with $z=0.959(54)$ with a 
chi-square per degree of freedom of $\chi^2/\mbox{\rm d.o.f.} = 1.04$ 
for the convex potential (CP), and with 
$z=0.808(42)$ 
with a $\chi^2$/d.o.f. 
of $0.58$ for the double-well potential (DW). These exponents are indeed well
comparable to the autocorrelation times for
the even moment $\la x^2 \ra$ which are $z=0.8356(92)$ (CP) and 
$z=0.715(27)$ (DW) \cite{js93a}.  

Also for the W-cycle the behaviour for the moments is qualitatively reproduced
by the virial estimator. For the average path $\la x \ra$ slowing down was 
completely eliminated with values of $z$ consistent with $0$, while for 
$\la x^2 \ra$ the exponents were found to be $z=0.1043(29)$ (CP) and 
$z=-0.015(11)$ (DW) \cite{js93a}.
For the virial estimator we find again an almost complete
reduction of critical slowing down with exponents of $z=0.128(12)$ with a 
$\chi^2$/d.o.f. of $0.20$ (CP) and $z= 0.0467(86)$ with a $\chi^2$/d.o.f. of 
$0.49$ (DW). The W-cycle thus almost completely eliminates slowing down in 
the continuum limit with absolute values of $\tau$ close to $0.5$ which 
means complete decorrelation in between measurements. 

Figures 2a and 2b show that the staging
algorithm also eliminates slowing down albeit with somewhat larger absolute
values for $\tau$. Judged from the exponents $z$ the staging algorithm
eliminates slowing down with exponents that are in fact perfectly consistent
with $0$ well within the statistical error bars. Here the fits give 
$z=0.008(17)$ with a $\chi^2$/d.o.f. of $0.86$ (CP) and $z=-0.005(20)$ with 
a $\chi^2$/d.o.f. of $0.33$ (DW).

Regarding the asymptotic behaviour the staging algorithm seems to be
slightly superior to the W-cycle where a small $L$-dependence cannot be
excluded from the data. One should also not forget that for one-dimensional 
systems the number of operations involved in the W-cycle scales with 
another $\log L$ dependence. For practical applications, however, it is also 
important to look at the absolute values of the autocorrelation times. These 
indeed turn out to be several times larger for the staging algorithm than 
for the W-cycle for grid sizes up to the largest one of $L=1024$
considered in our investigations. 
%
       \section{Discussion}\label{sect:Discussion}
%
Multigrid techniques and the staging algorithm provide two
different but equally successful path integral Monte Carlo methods.
An investigation of the integrated autocorrelation times
for the virial estimator of the internal energy shows that both the staging
algorithm and the multigrid schemes solve the slowing down problem of
local update schemes in the continuum limit.
For the staging algorithm this was demonstrated for the optimal choice of the 
parameter $j$ which determines the length of the staging segment.
This optimal choice differs notably from the one obtained following the common
rule of achieving a certain acceptance probability. It scales with the number
of lattice sites $L$ in the same way as does the correlation length along
the path measured in lattice units.

A comparison of the two update schemes from a practitioner's point of view
shows that they both have their advantages and drawbacks. 
The staging algorithm completely beats slowing down with exponents which are 
even smaller
than the ones for the multigrid W-cycle even though the absolute 
autocorrelation times are several times larger. One also has to take into 
account the number of operations per sweep which for one-dimensional systems 
grows proportional to $\log L$ for the W-cycle. Also in more technical respects
the staging algorithm is somewhat easier to implement for simple systems than
the recursive multigrid scheme. Even though this is obviously hardware
and platform dependent, one would probably find in most situations
similar to the one investigated here that the staging algorithm will be 
preferable
as far as CPU time is concerned. An advantage of the multigrid scheme is
the generality of its definition and the fact that it is mathematically well
defined and understood. Multigrid schemes are therefore quite easily 
generalizable to higher dimensions and quantum chains \cite{js95a}.
Also, they can readily be combined with reweighting techniques such as the 
multicanonical scheme useful for tunneling phenomena \cite{js94}.
Given the very good performance of the staging algorithm for simple systems
it would be interesting to work out similar generalizations for this update
scheme as well.
%
\section*{Acknowledgments}
%
W.J. thanks the Deutsche Forschungsgemeinschaft for a Heisenberg
fellowship.
%
               
%
\newpage
%
\clearpage
{\Large\bf Figure Captions}
%
  \vspace{1cm}
  \begin{description}
    \item[\tt\bf Fig. 1:] (a) Integrated autocorrelation times 
for the virial estimator and the convex potential
using the staging algorithm
for various grid sizes $L$.
The $x$-axis is the length of the staging segment.\\
(b) The same data as in (a), when plotted versus a rescaled $x$ variable, 
collapse onto a common master curve.
    \item[\tt\bf Fig. 2:] 
(a) Double logarithmic plot of integrated autocorrelation times $\tau$ vs 
the grid size $L$ for the virial estimator of the convex potential (CP) 
using the staging (S) algorithm and the multigrid V- and W-cycles. The straight
lines are fits to the data according to $\tau = \alpha L^z$, yielding
$z = 0.008(17)$ (S), $z = 0.959(54)$ (V), and $z = 0.128(12)$ (W).\\
(b) The same plot as in (a) for the double-well potential (DW). Here the fits
give $z = -0.005(20)$ (S), $z = 0.808(42)$ (V), and $z = 0.0467(86)$ (W).
\end{description}
\newpage
%
%
\begin{figure}[bhp]
\vskip 7.1truecm
\includegraphics{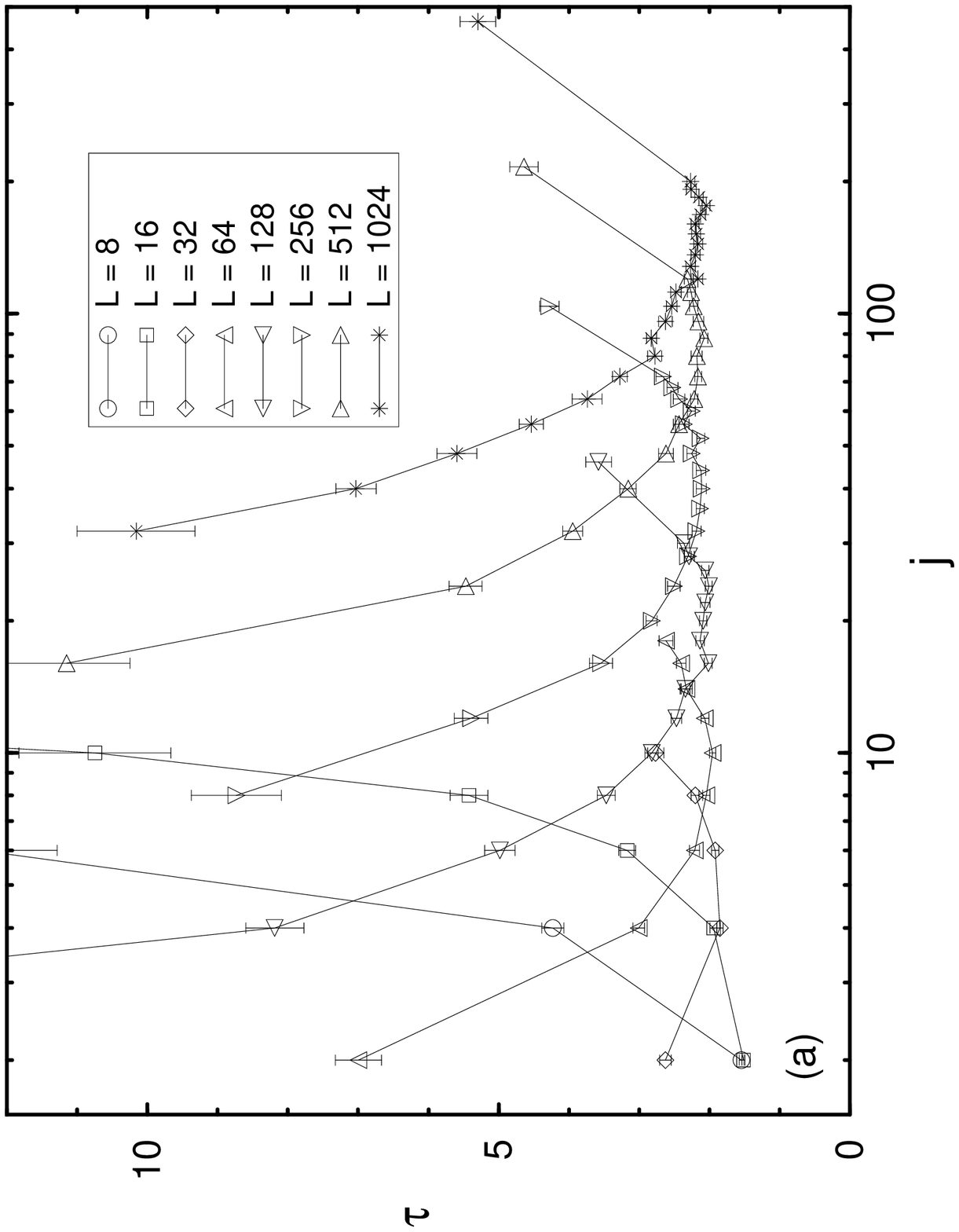}
\end{figure}
\begin{figure}[bhp]
\vskip 7.1truecm
\includegraphics{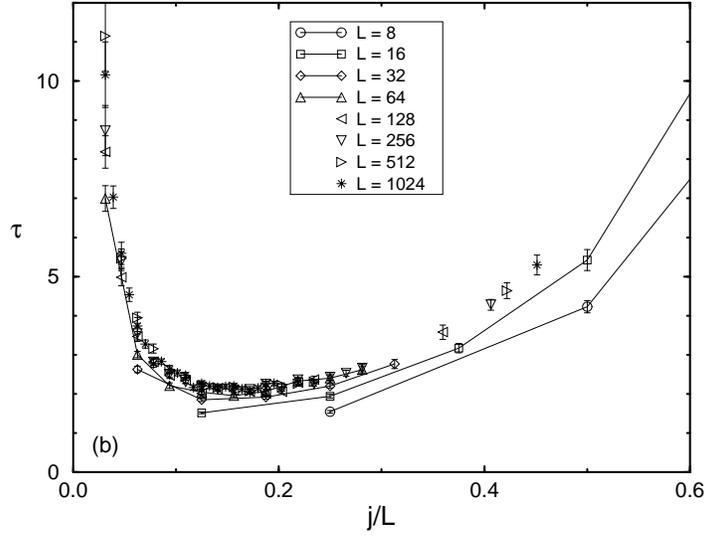}
\caption[a]{{
(a) Integrated autocorrelation times for the virial estimator and the 
convex potential using the staging algorithm for various grid sizes $L$.
The $x$-axis is the length of the staging segment.\\
(b) The same data as in (a), when plotted versus a rescaled $x$ variable,
collapse onto a common master curve.
}}
\end{figure}
\begin{figure}[bhp]
\vskip 7.1truecm
\includegraphics{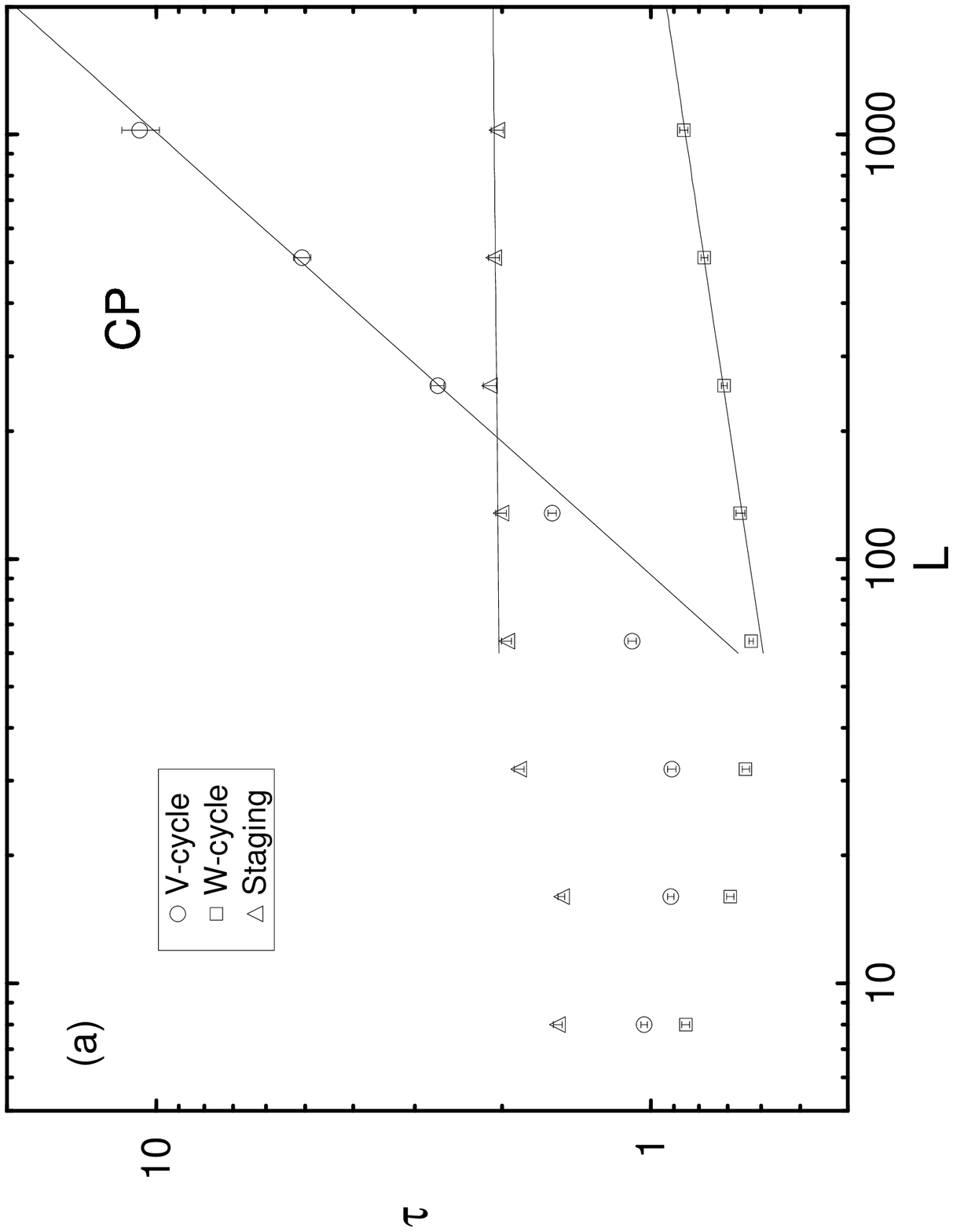}
\end{figure}
\begin{figure}[bhp]
\vskip 7.1truecm
\includegraphics{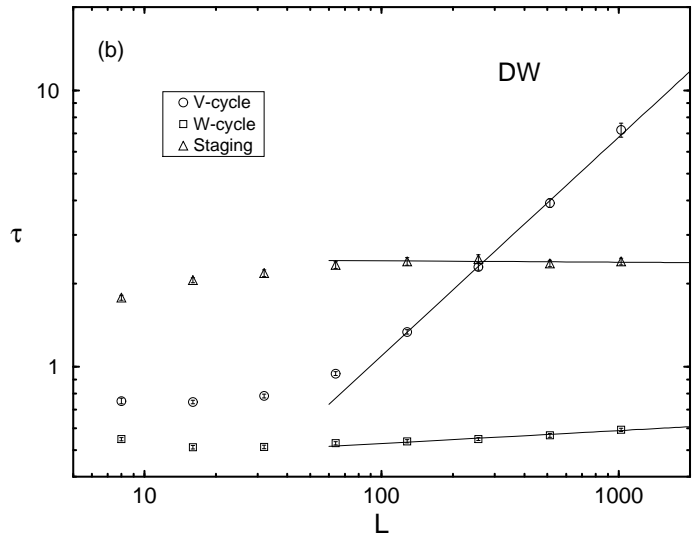}
\caption[a]{{ 
(a) Double logarithmic plot of integrated autocorrelation times $\tau$ vs
the grid size $L$ for the virial estimator of the convex potential (CP)
using the staging (S) algorithm and the multigrid V- and W-cycles. The straight
lines are fits to the data according to $\tau = \alpha L^z$, yielding
$z = 0.008(17)$ (S), $z = 0.959(54)$ (V), and $z = 0.128(12)$ (W).\\
(b) The same plot as in (a) for the double-well potential (DW). Here the fits
give $z = -0.005(20)$ (S), $z = 0.808(42)$ (V), and $z = 0.0467(86)$ (W).
}}
\end{figure}
\end{document}